\documentclass[aps,prl,twocolumn,showpacs]{revtex4}
\usepackage{graphicx}
\usepackage{dcolumn}   
\usepackage{bm}        
\usepackage{multirow}
\usepackage{amssymb}
\usepackage{amsfonts}   
\usepackage{amsmath} 
\usepackage{amsthm}
\usepackage{color}
\usepackage{marvosym}
\newcommand{\mathbbm}[1]{\text{\usefont{U}{bbm}{m}{n}#1}} 
\usepackage[utf8]{inputenc}

\newcommand{\tr}{\text{tr}}
\newcommand{\dyad}[1]{| #1\rangle \langle #1|}  
\newcommand{\expval}[2]{\langle #2 | #1 | #2 \rangle}
\newcommand{\conv}[1]{\operatorname{conv}(#1)}
\newcommand{\pd}[2]{\frac{\partial #1}{\partial #2}} 	
\newcommand{\argmin}[0]{\operatorname{argmin}}
\newcommand{\argmax}[0]{\operatorname{argmax}}
\newcommand{\mc}[1]{\mathcal{#1}}
\newcommand{\on}[1]{\operatorname{#1}}

\newcommand{\bra}[1]{\langle #1|}
\newcommand{\ket}[1]{|#1\rangle}

\newcommand{\ketbra}[2]{| #1\rangle \langle #2|}
\newcommand{\R}{\ensuremath{\mathbbm R}}

\newcommand{\be}{\begin{equation}}
\newcommand{\ee}{\end{equation}}
\newcommand{\bea}{\begin{eqnarray}}
\newcommand{\eea}{\end{eqnarray}}

\newcommand{\eins}{\mathbbm{1}}
\newcommand{\WW}{\ensuremath{\mathcal{W}}}

\newcommand{\MM}{\ensuremath{\mathcal{M}}}
\newcommand{\HH}{\ensuremath{\mathcal{H}}}
\newcommand{\FF}{\ensuremath{\mathcal{F}}}

\newcommand{\QQ}{\ensuremath{\mathcal{Q}}}
\newcommand{\GG}{\ensuremath{\mathcal{G}}}

\newcommand{\kommentar}[1]{}

\renewcommand{\vr}{\ensuremath{\varrho}}

\newcommand{\forget}[1]{}

\begin{document}
\title{Characterizing ground and thermal states of few-body Hamiltonians}
\date{\today}
\author{Felix Huber}
\author{Otfried G\"uhne}
\affiliation{Naturwissenschaftlich-Technische Fakult\"at, 
Universit\"at Siegen, Walter-Flex-Str. 3, 57068 Siegen, Germany}

\begin{abstract}
The question whether a given quantum state is a ground or thermal state 
of a few-body Hamiltonian can be used to characterize the complexity of 
the state and is important for possible experimental implementations. 
We provide methods to characterize the states generated by two- and, 
more generally, $k$-body Hamiltonians as well as the convex hull of 
these sets. This leads to new insights into
the question which states are uniquely determined by their marginals and
to a generalization of the concept of entanglement. Finally, certification
methods for quantum simulation can be derived.
\end{abstract}

\pacs{03.65.Ud, 03.67.Mn}

\maketitle

{\it Introduction.---}
Interactions in quantum mechanics are described by Hamilton operators. 
The study of their properties, such as their symmetries, eigenvalues, 
and ground states, is  central for several fields of physics. 
Physically relevant Hamiltonians, however, are often restricted to 
few-body interactions, as the relevant interaction mechanisms are 
local. But the characterization of generic few-body Hamiltonians 
is not well explored, since in most cases one starts with a 
given Hamiltonian and tries to find out its properties.

In quantum information processing, ground and thermal states of local 
Hamiltonians are of interest for several reasons: First, if a desired 
state is the ground or thermal state of a sufficiently local Hamiltonian, it might 
be experimentally prepared by engineering the required interactions
and cooling down or letting thermalise the physical system \cite{general}. 
For example, one may try to prepare a cluster state, the resource for measurement-based quantum 
computation, as a ground  state of a local Hamiltonian \cite{VandenNest2008}. 
Second, on a more theoretical side, ground states of $k$-body Hamiltonians 
are completely characterized by their reduced $k$-body density matrices. 
The question which states are uniquely determined by their marginals 
has been repeatedly studied and is a variation of the representability 
problem, which asks whether given marginals can be represented by a 
global state \cite{marginalreview}. It has turned out that many pure states have the 
property to be uniquely determined by a small set of their marginals 
\cite{Linden2002, jones}, and for practical purposes it is relevant that often entanglement 
or non-locality can be inferred by considering the marginals only \cite{marginalentanglement}. 

In this paper we present a general approach to characterize ground
and thermal states of few-body Hamiltonians.  We use the formalism 
of exponential families, a concept first introduced for classical 
probability distributions by Amari~\cite{Amari2001} and extended to 
the quantum setting in Refs.~\cite{Hasegawa1997, Zhou08, Zhou09, Niekamp13}. 
This offers a systematic characterization of 
the complexity of quantum states in a conceptionally pleasing way.
We derive two methods that can be used to compute various distances to
thermal states of $k$-body Hamiltonians: The first method is general and
uses semidefinite programming, while the second method is especially 
tailored to cluster and, more generally, graph states. In previous 
approaches it was only shown that some special states are far away from
the eigenstates of local Hamiltonians \cite{Haselgrove2003}, but no
general method for estimating the distance is known.

Our approach leads to new insights in various directions. First, it has been 
shown that cluster and graph states can, in general, not be exact 
ground states of two-body Hamiltonians \cite{VandenNest2008}, 
but it was unclear whether they still can be approximated sufficiently well.
Our method shows that this is not the case and allows to bound the distance
to ground and thermal states. Second, as shown in Ref.~\cite{Linden2002}, almost all 
pure states of three qubits are completely determined by their two-party 
reduced density matrices. As we prove, for \(N\geq5\) qubits or four qutrits
this is not the case, but we present some evidence that the fact might
still be true for four qubits. Finally, our method results in witnesses, 
which can be used in a quantum simulation experiment to certify that a 
three-body Hamiltonian or a Hamiltonian having long-range interactions 
was generated.

\begin{figure}[t]
\includegraphics[width=0.85\columnwidth]{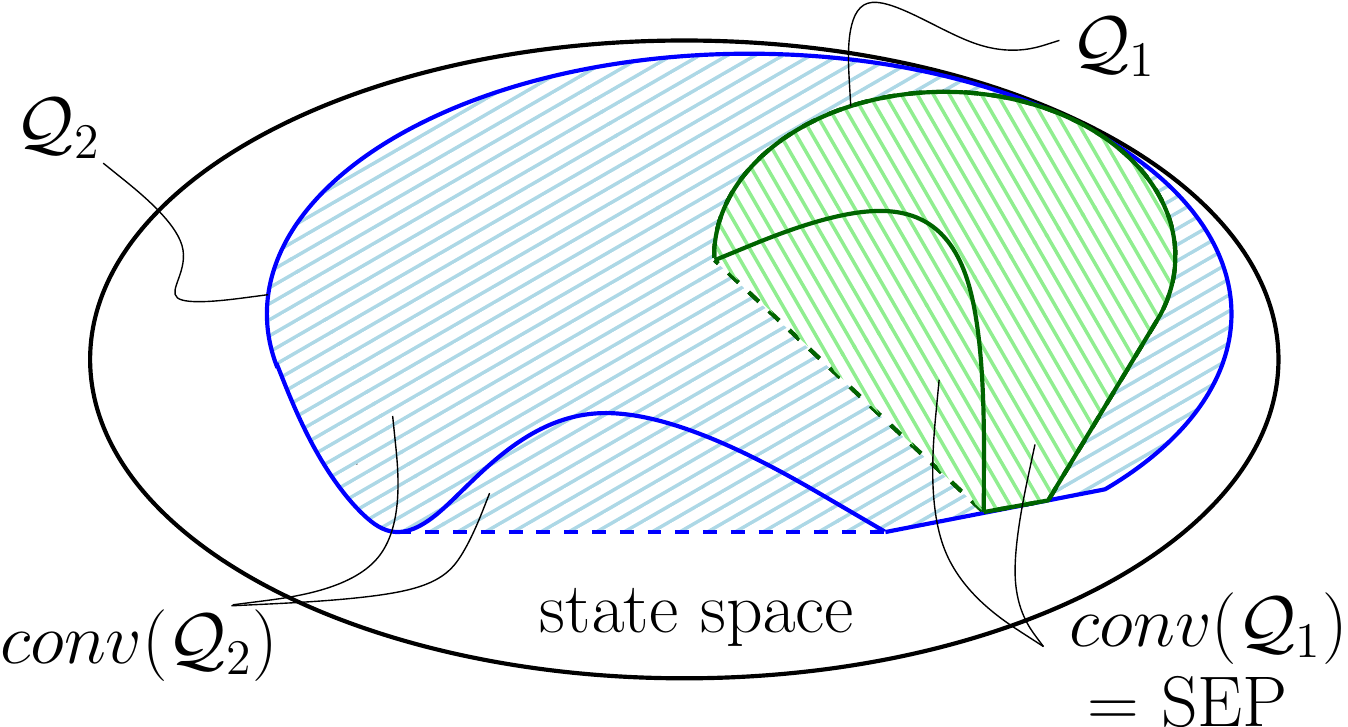}
\caption{Schematic view of the state space, the exponential 
families $\QQ_1$ and $\QQ_2$, and their convex hulls. While the whole space
of mixed states is convex, the exponential families are non-convex 
low-dimensional manifolds. The convex hull of $\QQ_1$ are the fully separable states
and our approach allows to characterize the convex hull for arbitrary $\QQ_k$.
}
\label{fig:convhull}
\end{figure}

{\it The setting.---}
A two-local (or two-body) Hamiltonian of a system consisting of 
$N$ spin-1/2 particles can be written as 
\begin{equation} 
\label{eq:2localHamiltonian}
H = 
\sum\nolimits_{i,j=1}^{N} \sum\nolimits_{\alpha \beta}
\lambda_{\alpha\beta}^{(ij)} 
\sigma_\alpha^{(i)}\otimes\sigma_\beta^{(j)}\,,
\end{equation}
where $\sigma_\alpha^{(i)}$ denotes a Pauli matrix 
$\{\eins, \sigma_x, \sigma_y, \sigma_z\}$ 
acting on the $i$-th particle etc. Note that the identity matrix
is included, so $H$ can also contain single particle terms.
We denote the set of all possible two-local Hamiltonians 
by \(\mc{H}_2\) and in an analogous manner the set 
of $k$-local Hamiltonians by $\HH_k.$ 
An example for a two-local Hamiltonian is
the Heisenberg model having nearest-neighbor interactions. 
However, our approach generally ignores any geometrical arrangement 
of the particles.
Finally, for an arbitrary  multi-qubit operator $A$ we 
call the number of qubits where it acts on non-trivially the {\it weight} of $A$.
In practice, this can be determined by expanding $A$ in terms of 
tensor products of Pauli operators and looking for the largest 
non-trivial product.

The set we aim to characterize is the so-called exponential family 
\(\QQ_2\), consisting of thermal states of two-local Hamiltonians
\begin{equation}
\QQ_2 = \big\{ \tau \,\big\vert \, 
\tau = \frac{e^{-\beta H}}{\tr[e^{-\beta H}]}\,, H \in \mc{H}_2 \big\}\,.
\end{equation}
Ground states can be reached in the limit of infinite inverse temperature
\(\beta\). For any \(k\), the exponential families \(\QQ_k\) can be 
defined in a similar fashion. The set \(\QQ_1\) consists of mixed 
product states, the set \(\QQ_N\) of the full state space. The 
exponential families form the hierarchy 
\(\QQ_1 \subseteq \QQ_2 \subseteq \cdots \subseteq  \QQ_N\), 
and a suitable \(\beta H\) can be seen as a way of parameterizing a 
specific density matrix \( \tau =  e^{-\beta H} / \tr[e^{-\beta H}]\).
The question arises, what states are in \(\QQ_k\)? And for those 
which are not, what is their best approximation by states in 
\(\QQ_k\)?

It turns out to be fruitful to consider the convex hull
\begin{equation}
  \conv{\QQ_2} = 
  \big\{ 
  \sum\nolimits_i p_i \, \tau_i \, \, | \,\, \tau_i \in \QQ_2\,,\,
  \sum\nolimits_i p_i = 1 \, , \, p_i \geq 0 \big\}\nonumber\,,
  \end{equation}
and ask whether a state is in this convex hull or not (see 
also Fig.~\ref{fig:convhull}). The convex hull has a clear physical 
interpretation as it contains all states that can be generated 
by preparing thermal states of two-body Hamiltonians stochastically with 
probabilities $p_i$. In this way,  taking the convex hull can 
be seen as a natural extension of the concept of entanglement: 
The thermal states of one-body Hamiltonians are just the mixed 
product states and their convex hull are the fully separable 
states of $N$ particles \cite{Guehne09}. 
In this framework,
the result of Linden et al.~\cite{Linden2002} can be rephrased as 
stating that all three-qubit states are in the closure of the convex 
hull \(\conv{\QQ_2}\), since nearly all pure states are ground 
states of two-body Hamiltonians. 

Finally, the characterization of the convex hull leads to the concept of 
witnesses that can be used for the \emph{experimental} detection of 
correlations \cite{Guehne09}. Witnesses are observables which have positive 
expectation values for states inside a given convex set. Consequently, the observation
of a negative expectation value proves that a state is outside of the set. 
We will see below that such witnesses can be used to certify quantum simulation.

{\it Quantum exponential families.---} 
We recall some results on the characterization 
of quantum exponential families \cite{Zhou09, Niekamp13}. Given a state \(\varrho\), consider 
its distance from the exponential family \(\QQ_2\) in terms of the relative 
entropy (or divergence) $S(\varrho || \tau)=\tr[\varrho (\log(\varrho)-\log(\tau))]$. 
As the closest state to \(\varrho\) in \(\QQ_2\), one obtains the 
so-called information projection \(\tilde{\varrho}_2\). It has 
been shown that the following three characterizations for the 
information projection \(\tilde{\varrho}_2 \in \QQ_2\) are equivalent 
\cite{Zhou09}:

\noindent
(a) \(\tilde{\varrho}_2\) is the unique minimizer of the relative entropy of 
\(\varrho\) from the set \(\QQ_2\), 
\begin{equation}
\tilde{\varrho}_2 = \argmin_{\tau \in\QQ_2} S(\varrho || \tau) \, .
\end{equation}
(b) Of the set of states having the same two-body reduced density matrices ($2$-RDMs)
as \(\varrho\), denoted by \(\mc{M}_2(\varrho)\), \(\tilde{\varrho}_2\) has a maximal 
von Neumann entropy
\begin{equation}
  \tilde{\varrho}_2 = \argmax_{\mu\in\mc{M}_2(\varrho)} S(\mu) \,. \label{eq:exp_family2}
\end{equation}
(c) Finally, \(\tilde{\varrho}_2\) is the unique intersection of \(\QQ_2\) and 
\(\mc{M}_2(\varrho)\). 
{From} (b)  it  follows that if for a state \(\sigma\) another state \(\varrho\) 
of higher entropy but having the same \(2\)-RDMs can be found, then \(\sigma\) must 
lie outside of \(\QQ_2\). A further discussion can be found in Appendix A~\cite{suppmat}.

States not in \(\QQ_2\) are said to have irreducible correlations of
order three or higher, because they contain information which is not already 
present in their \(2\)-RDMs, if one wishes to reconstruct the global state 
from its marginals according to Jaynes' maximum entropy principle \cite{Teo2011}. This is 
conceptionally nice, but also has certain drawbacks. Importantly, the irreducible 
correlation as quantified by the relative entropy is not continuous, as shown 
in Ref.~\cite{Weis2012}. In addition, the relative entropy 
is difficult to estimate experimentally without doing
state reconstruction, so other distances such as the fidelity are preferable.
These properties make the relative entropy somewhat problematic
and give further reasons why we consider the convex hull.

{\it Characterization via semidefinite programming.---}
Our first method to estimate the distance of a given state to the convex hull
of  \(\QQ_2\) relies on semidefinite programming \cite{Boyd}. This optimization method
is insofar useful, as semidefinite programs are efficiently solvable
and their solutions can be certified to be optimal. Moreover, 
ready-to-use packages for their implementation are available.

As a first step we formulate a semidefinite program to test 
if a given pure \(\ket{\psi}\) state is outside of \(\QQ_2\).  
{From} the characterization in Eq.~\eqref{eq:exp_family2} it follows 
that it suffices to find a different state \(\varrho\) having the 
same \(2\)-RDMs as \(\ket{\psi}\). If \(\varrho\) is mixed, its 
entropy is higher than that of \(\ket{\psi}\), meaning that 
$\ket{\psi}$ cannot be its own information projection and therefore lies outside of
$\QQ_2$. If \(\varrho\) is pure, consider the convex combination 
\((\dyad{\psi} + \varrho)/2\), again having a higher entropy. 
To simplify notation we define for an arbitrary \(N\)-qubit operator \(X\) the operator 
\(R_k(X)\) as the projection of \(X\) onto those operators, which can be decomposed into terms 
having at most weight \(k\).
In practice, \(R_k(X)\) can be 
computed by expanding \(X\) in Pauli matrices, and removing all 
terms of weight larger than \(k\). Note that \(R_k(\vr)\) may 
have negative eigenvalues.

The following semidefinite program finds a state with the same
$k$-body marginals as a given state \(\ket{\psi}\):
\begin{align}
  \min_\vr: &\quad \tr [\varrho \dyad{\psi}]  \nonumber\\
  \text{subject to:}&\quad R_k(\varrho) = R_k(\dyad{\psi}),    \nonumber\\
 &\quad \tr[\varrho] = 1, \quad \varrho = \varrho^{\dag}, \quad \varrho \geq \delta\eins \,.
 \label{sdptest}
\end{align}
While this program can be run with 
\(\delta=0\), it is useful to choose \(\delta\) to be 
strictly positive. Then, a strictly positive $\vr$ may be found, which 
is guaranteed to be distant from the state space boundary. Consequently, 
if $\ket{\psi}$ is disturbed, one can still expect to find a state with the 
same reduced density matrices in the vicinity of $\vr$. This can be
used to prove that the distance to $\QQ_2$ is finite, and will 
allow us to construct witnesses for proving irreducible correlations
in $\ket{\psi}$.
We make this rigorous in the following Observation. 
For that, let \(\mc{B}(\ket{\psi})\) be the 
ball in trace distance \(D_{\tr}(\mu, \eta) = \frac{1}{2}\tr(|\mu - \eta|)\) 
centered at \(\ket{\psi}\).

{\bf Observation 1.}
{\it Consider a pure state \(\ket{\psi}\) and a mixed state 
  \(\varrho \geq \delta\eins\) with \(R_k(\varrho)= R_k(\dyad{\psi})\). 
  Then, for any state \(\sigma\) in the ball \(\mc{B}_\delta(\ket{\psi})\) 
  a valid state \(\tilde{\varrho}\) in \(\mc{B}_\delta(\varrho)\) can be 
  found, such that their \(k\)-party reduced density matrices match.
  Moreover, the entropy of $\tilde\vr$ is larger than or equal to the entropy of $\sigma$.
  This implies that the ball \(\mc{B}_\delta(\ket{\psi})\) contains no
  thermal states of $k$-body Hamiltonians.
}
  
The proof is given in Appendix B \cite{suppmat}.

In the Observation, we considered the trace distance, but a ball in fidelity instead 
of trace distance can be obtained: Consider a state $\sigma$ near \(\ket{\psi}\), 
having the fidelity \(F(\sigma, \psi) = \alpha \geq 1 - \delta^2\), 
where \(F(\varrho, \psi) = \tr[\varrho \dyad{\psi}] = \bra{\psi} \varrho 
\ket{\psi}\). Then from the Fuchs-van-de-Graaf 
inequality follows 
\(
D_{tr}(\sigma, \dyad{\psi}) 
\leq
\sqrt{1-F(\sigma, {\psi})} 
\leq  \delta \, ,
\)
and Observation~1 is applicable \cite{fuchs}. 

The usage of the fidelity as a distance measure has a clear 
advantage from the experimental point of view, as it allows the construction of witnesses for 
multiparticle correlations. Indeed the observable
\begin{equation}
\WW = (1-\delta^2) \eins - \dyad{\psi}
\end{equation}
has a positive expectation value on all states in $\QQ_k$ and, due to the linearity
of the fidelity, also on all states within the convex hull ${\rm conv}(\QQ_k)$.
So, a negative expectation value signals the presence of $k$-body correlations.
Witnesses for entanglement have already found widespread applications in experiments
\cite{Guehne09}.

Equipped with a method to test whether a pure state is in \(\conv{\QQ_2}\) 
or not we are able to tackle the question whether the results of Ref.~\cite{Linden2002}
can be generalized. Recall that in this reference it has been shown that nearly
all pure states of three qubits are uniquely determined (among all mixed states)
by their reduced two-body density matrices. This means that they are ground
states of two-body Hamiltonians. Consequently, the closure of the convex hull
\(\conv{\QQ_2}\) contains all pure states and therefore also all mixed states, 
and the semidefinite program in Eq.~(\ref{sdptest}) will not be feasible for
$\delta$ strictly positive. The question is whether this result holds for more qubits too.

Concerning pure five-qubit states, we numerically found a fraction of \(40\%\) 
to be outside of \(\conv{\QQ_2}\). 
In the case of pure four-qubit states however, no tested random state has been 
found to lie outside of \(\conv{\QQ_2}\). 
Given the fact that the test works well in the cases of five and six qubits,
this leads us to conjecture that nearly all pure four qubit states are 
in \(\conv{\QQ_2}\), and hence also in \(\QQ_2\).
This would imply that a similar result as the one obtained by 
Ref.~\cite{Linden2002} holds in the case of four qubits:
almost every pure state of four qubits is completely determined by its two-particle 
reduced density matrix. More details are given in Appendix C \cite{suppmat}.

\begin{figure}[t]
\includegraphics[width=0.85 \columnwidth]{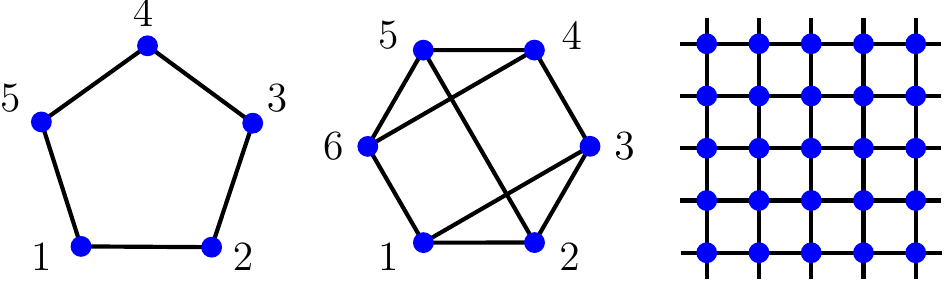}
\caption{Examples of graphs discussed in this paper. 
Left: The five-qubit ring-cluster graph. The corresponding ring-cluster 
state \(\ket{C_5}\) has a finite distance to the exponential family 
$\QQ_2$.
Middle: The maximally entangled six-qubit $\ket{M_6}$ state is not in 
the convex hull of $\QQ_3$.
Right: The \(2\)D periodic \(5\times5\) cluster state \(\ket{C_{5\times5}}\) 
is not in \(\conv{\QQ_4}\).}
  \label{fig:ringcluster5}
\end{figure}

{\it Characterization via the graph state formalism.---}
The family of graph states includes 
cluster states and GHZ states and has turned out to be
important for measurement-based quantum computation and
quantum error correction \cite{hein}. Due to their importance, the question
whether graph states can be prepared as ground states of two-body 
Hamiltonians has been discussed before \cite{VandenNest2008}. Generally, graph states 
have shown to not be obtainable as unique non-degenerate ground states 
of two-local Hamiltonians. Further, any ground state of a $k$-local 
Hamiltonian \(H\) can only be \(\epsilon\)-close to a graph state 
\(\ket{G}\) with \(m(\ket{G}) > k\) at the cost of \(H\) 
having an \(\epsilon\)-small energy gap relative to the total energy 
in the system \cite{VandenNest2008}. Here \(m(\ket{G})\) is the minimal 
weight of any element in the stabilizer \(S\) of state \(\ket{G}\) (see also
below). But as pointed out in Ref.~\cite{VandenNest2008}, 
this does not imply that graph states cannot be approximated in 
general, as \(\epsilon\) is a relative gap only.

Let us introduce some facts 
about graph states. A graph  consists of vertices and edges 
(see Fig.~\ref{fig:ringcluster5}).
This defines the generators 
\begin{equation}
g_a = \sigma_x^{(a)} \prod\nolimits_{ b\in N(a)} \sigma_z^{(b)},
\end{equation}
where the product of the $\sigma_z^{(b)}$ runs over all 
vertices connected to vertex \(a\), called neighborhood 
\(N(a)\). The graph state $\ket{G}$ can be defined as the unique 
eigenstate of all the $g_a$, that is $\ket{G}=g_a \ket{G}$. This
can be rewritten with the help of the stabilizer. The stabilizer 
\(S\) is the commutative group consisting of all possible $2^N$ products 
of $g_a$, that is \(S=\{s_i=\prod_{a\in I}g_a\}\). Then, the graph state can be written 
as
\(
  \dyad{G} = 2^{-N} \sum_{s_i \in S} s_i
\)  \cite{hein}.
This formula allows to determine the reduced density matrices 
of graph states easily, since one only has to look at the products of the 
generators $g_a$. 

For instance, all stabilizer elements of the five-qubit ring cluster state \(\ket{C_5}\) have at 
least weight three, 
and therefore the \(2\)-RDMs of \(\ket{C_5}\) are maximally mixed.
By choosing \(\delta = 2^{-5}\) in Observation \(1\), the maximum overlap to \(\conv{\QQ_2}\) is bounded
by \(F_{\tau \in \QQ_2}(\ket{C_5}, \tau) \leq 1 - \delta^2 \approx 0.99902\).
Note that Ref.~\cite{NiekampThesis} has demonstrated a 
slightly better bound \(F(\ket{C_5}, \tau) \leq 1/32+\sqrt{899/960} \approx 0.99896\). 
However, both bounds are by far not reachable in current experiments. 
In fact, one can do significantly better. 
In the following, we will formulate 
a stricter bound by first considering \(\QQ_2\) and the ring cluster state \(\ket{C_N}\)
for an arbitrary number of qubits $N\geq5$, but the result is general.

{\bf Observation 2.}
{\it 
  The maximum overlap between the $N$-qubit ring cluster state 
  \(\ket{C_N}\) and an \(N\)-qubit state \(\tau \in \QQ_2\) 
  is bounded by
  \begin{equation} 
  \label{theorem:overlap}
    \sup_{\tau \in \QQ_2} \expval{\tau}{C_N} \leq \frac{D-1}{D} \,,
  \end{equation}
  where $D=2^N$ is the dimension of the system. More generally, for an arbitrary pure 
  state with maximally mixed reduced $k$-party states in a $d^{\otimes{N}}$-system, 
  the overlap with $\QQ_k$ is bounded by $(d^N-1)/d^N$. 
}
  
The proof is given in Appendix D \cite{suppmat}.  
  
In the case of five qubits, \(F_{\tau \in \QQ_2}(\ket{C_5}, \tau) \leq 31/32 \approx 0.96875\),
which improves the bound on the distance to conv$(\QQ_2)$ by more than two
orders of magnitude \cite{fnx}.
From Observation 2, we can construct the witness 
\begin{equation}
\WW = \frac{D-1}{D} \eins - \dyad{C_N}\,,
\end{equation}
which detects states outside of \(\conv{\QQ_2}\). In a similar fashion,  
any state having the maximally mixed state as $k$-particle RDMs can be 
used to construct a witness for \(\conv{\QQ_k}\). First,
there is a four-qutrit state with maximally mixed 2-RDMs \cite{goyeneche}, 
which can be used to derive a 
witness for \(\conv{\QQ_2}\). The highly entangled six-qubit state 
$\ket{M_6}$ (see the graph in Fig.~\ref{fig:ringcluster5}) has maximally 
mixed 3-RDMs, so $\WW = \frac{63}{64} \eins - \dyad{M_6}$ is a witness to exclude thermal 
states of three-body Hamiltonians. Third, consider 
a $5\times 5$ \(2\)D cluster state with periodic boundary conditions (see 
Fig.~\ref{fig:ringcluster5}). This state has \(m(\ket{C_{5 \times 5}})=5\) \cite{VandenNest2008},
and can therefore serve as a witness \(\mc{W} = \alpha\eins - \dyad{C_{5 \times 5}}\)
for \(\conv{\QQ_4}\), where \(\alpha = (2^{25}-1) / 2^{25}\).
It should be noted that this witness can also be used for \(\conv{\QQ_2}\), for which the value \(\alpha\) might be
improved \cite{h4note}.
Finally, the minimal distance \(D_k\) in terms of the relative entropy from 
\(\QQ_k\) can be lower bounded by the fidelity distance from 
its convex hull \(\conv{\QQ_k}\), see Appendix D for details \cite{suppmat}.

{\it Quantum simulation as an application.---}
The aim of quantum simulation is to simulate a physical system of 
interest by another well-controllable one. Naturally, it is crucial 
to ascertain that the interactions really perform as intended. 
Different proposals have recently come forward to engineer sizeable three-body 
interactions in systems of cold polar molecules \cite{Buechler2007}, 
trapped ions \cite{Bermudez09}, ultracold atoms in triangular lattices \cite{Pachos2004}, Rydberg atoms \cite{Weimer2010} and circuit QED 
systems \cite{Hafezi2013}.
Using the ring cluster state witness \(\mc{W} = \alpha\eins - \dyad{C_N}\) 
derived above, it is possible to certify that three- or higher-body interactions 
have been engineered. This is done by letting the system under control thermalise. 
If then \(\langle\mc{W}\rangle < 0\) is measured, one has 
certified that interactions of weight three or higher are present. 
At least five qubits are generally required for this, but by further restricting the 
interaction structure, four qubits can be enough for demonstration purposes.
This can already be done with a fidelity of \(93.75\%\), 
which is within reach of current technologies.
Further details can be found in Appendix E \cite{suppmat}.

As an outlook, one may try to extend this idea of interaction certification 
to the unitary time evolution under local Hamiltonians. For instance, 
digital quantum simulation can efficiently approximate the time evolution 
of a time-independent local Hamiltonian and in Ref.~\cite{Lanyon07102011} 
an effective \(6\)-particle interaction has been engineered by applying 
a stroboscopic sequence of universal quantum gates. The process fidelity 
was quantified using quantum process tomography, however it would be of 
interest to prove that the same time evolution cannot be generated by 
\(5\)-particle interactions only.

{\it Conclusion.---} We have provided methods to characterize thermal and ground states 
of few-body Hamiltonians. Our results can be used to test experimentally
whether three-body or higher-order interactions are present. For future work, it 
would be desirable to characterize the entanglement properties of $\QQ_2$, e.g.
to determine whether the entanglement in these states is bounded, or whether they
can be simulated classically in an efficient manner.  Furthermore, it is of
significant experimental relevance to develop schemes to certify that a unitary
time evolution was generated by a $k$-body Hamiltonian.

{\it Acknowledgement.---}
We thank Tobias Galla, Sönke Niekamp, and Marcin Paw{\l}owski
for discussions. 
This work was supported by the SNSF, 
the COST Action MP1209, 
the FQXi Fund (Silicon Valley Community Foundation), 
the DFG, and the ERC (Consolidator Grant 683107/TempoQ).

\section{Appendix}

\subsection{A. Further discussion of the marginal set \(\MM_k(\varrho)\) and its relation to \(\QQ_k\)}
The marginal set \(\MM_k(\varrho)\) consists of quantum states having the same \(k\)-party reduced density matrices (\(k\)-RDMs) as \(\varrho\)
\begin{equation}
  \MM_k(\varrho) = \{ \mu \,|\, \mu_A = \varrho_A \text{ for all } |A| \leq k\} \,,
\end{equation}
where  \(\mu_A\) is the reduced state obtained by tracing out all subsystems not contained in \(A\).
This set is convex, as its states stay in the marginal family under convex combination.

The exponential family \(\QQ_k\) consists of thermal states of \(k\)-local Hamiltonians
\begin{equation}
\QQ_2 = \big\{ \tau \,\big\vert \, 
\tau = \frac{e^{H}}{\tr[e^{H}]}\,, H \in \mc{H}_2 \big\}\,.
\end{equation}
In contrast to the marginal set, the exponential families \(\QQ_k\) are, apart from \(\QQ_n\), not convex. To see this, note that \(\conv{\QQ_1}\) is the set of separable states, having a volume and a number of free parameters corresponding to the dimension of the state space. In addition, \(\conv{\QQ_k}\) is larger than the set of separable states for \( k \geq 2\). 
However, \(\QQ_k\) has not as many free parameters and is a set of measure zero. Thus \(\QQ_k \subsetneq \conv{\QQ_k}\), and \(\QQ_k\) cannot be convex.

The relations between the marginal set and the exponential family originates in two special ways to parametrize a quantum state \cite{Hasegawa1997}. These are the affine (also called mixed) and the exponential representations
\begin{align}
  \varrho_\text{aff} &= \eins / D + \eta_i A_i \,,		&\eta&\in (-1,1)^{D^2 - 1}  \,,  \nonumber\\
  \varrho_{\exp}     &= \exp[\theta_i A_i - \psi(\theta)] \,, 	&\theta &\in \R^{D^2 - 1}  \,,
\end{align}
where \(D\) is the dimension of the system, \(\{A_i\}\) is a suitable orthonormal basis of the operator space, and we sum over repeated indices. The Massieu function \(\psi(\theta) = \log \tr [\exp H]\) is not only required for normalization, but also defines, together with the potential \(\phi(\eta) = -S(\eta) = \tr [\varrho \log \varrho]\), a Legendre transform 
\(
  \psi(\theta) + \phi(\eta) - \theta_i \eta_i = 0\,.
\)
The relations
\begin{equation}
  \eta^i  = \frac{\partial \psi(\theta)}{\partial \theta_i} = \tr [\varrho A_i]\,,  \quad \theta_i  = \frac{\partial \phi(\eta)}{\partial \eta_i} = \tr[H A_i]\,,
\end{equation}
follow.
For any two states \(\varrho(\eta)\) and \(\varrho'(\theta')\), the following Pythagorean relation for the relative entropy holds
\begin{equation}
  S(\varrho || \varrho') = \phi(\eta) + \psi(\theta') -  \eta_i \theta'_i \,,
\end{equation}
and its repeated application yields
\begin{align}
  &S(\varrho || \varrho'') \nonumber\\
  &= S(\varrho || \varrho') + S(\varrho'|| \varrho'') +  (\eta_i - \eta'_i) \cdot (\theta'_i - \theta''_i) \,.
\end{align}

\begin{figure}[t]
\includegraphics[width=0.6 \columnwidth]{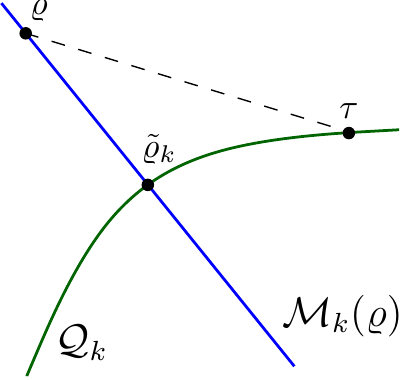}
\caption{The information projection \(\tilde{\varrho}_k\) lies in the unique intersection of \(\QQ_k\) and \(\MM_k(\varrho)\). It is also the minimizer of the relative entropy \(S(\rho || \cdot) \) in \(\QQ_k\).}
  \label{fig:information_projection}
  \end{figure}

The information projection \(\tilde{\varrho}_k\) of \(\varrho\) is the element in \(\MM_k(\varrho)\) having the largest von Neumann entropy.
Given \(\varrho\), its information projection \(\tilde{\varrho}_k\), and a \(\tau \in \QQ_k\), the Pythagorean relation then simplifies to
\begin{equation}
  S(\varrho || \tau) = S(\varrho || \tilde{\rho}_k) + S(\tilde{\varrho}_k || \tau)\,.
\end{equation}
The above definition of the information projection is equivalent to \(\tilde{\varrho}\) being in the unique intersection of the exponential family \(\QQ_k\) with \(\MM_k(\varrho)\), and to \(\tilde{\varrho}_k = \argmin_{\tau \in \QQ_k} S(\rho || \tau)\). This is illustrated in Fig.~\ref{fig:information_projection}.

We provide two examples. First, consider the five-qubit ring cluster state \(\ket{C_5}\) and its information projection onto \(\QQ_2\). The state \(\ket{C_5}\) has maximally mixed \(2\)-body marginals, and of the set \(\MM_2(\ket{C_5})\), the maximally mixed state has the highest entropy. 
Second, consider the one-parameter family of states 
\(\ket{GHZ_\alpha} = (\ket{000} + e^{i\alpha}\ket{111})/\sqrt{2} \). All of its two-party reduced states are equal to \((\dyad{00} + \dyad{11})/2\). 
Also,
\begin{align}
  \gamma &= \int_\alpha \! \dyad{GHZ_\alpha} \,\mathrm{d}\alpha \nonumber\\
  &= (\ketbra{000}{000} + \ketbra{111}{111}) / 2
\end{align}
has the same \(2\)-RDMs and is thus an element of the marginal set \(\MM_2(\ket{GHZ_\alpha})\). Additionally, \(\gamma\) is the information projection of \(\ket{GHZ_\alpha}\) onto \(\QQ_2\), as it is the element of maximum entropy in \(\MM_2(\ket{GHZ_\alpha})\). 
As known from Ref.~\cite{Linden2002}, almost all three qubit states are determined by their 2-RDMs, and thus the irreducible three-body correlation is discontinuous at \(\ket{GHZ_\alpha}\). More examples can be found in Ref.~\cite{Niekamp13}.

As a last part in this section, we relate the exponential family \(\conv{\QQ_k}\) to the sets of ground and excited states of local Hamiltonians respectively. As argued above, nondegenerate ground states of k-local Hamiltonians \(H_k \in \HH_k\) are determined by their k-RDMs and belong to the closure of \(\QQ_k\). Nondegenerate excited states of \(k\)-local Hamiltonians are completely determined by their \(2k\)-RMDs \cite{Chen2012}, and are therefore ground states of suitable \(2k\)-local Hamiltonians \(H_{2k} \in \HH_{2k}\).
The argument rests on the fact that any eigenstate of a Hamiltonian \(H_k\) will also be the ground state of \( (H_k - \lambda \eins)^2 \), where \(\lambda\) is the corresponding eigenvalue. A similar argument also holds for nondegenerate ground and eigenstates.
But as can be seen by parameter counting, there exist \(2k\)-local Hamiltonians which cannot be written as \(H_{2k} = (H_k - \lambda \eins)^2\) with \(H_k \in \HH_k\). Thus the set of eigenstates of \(k\)-local Hamiltonians \(\on{ES}(\HH_k)\) is a proper subset of the set of ground states of \(2k\)-local Hamiltonians \(\on{GS}(\HH_{2k})\), \(\on{ES}(\HH_k) \subsetneq \on{GS}(\HH_{2k}) \). 
Finally, thermal states of \(k\)-local 
Hamiltonians are in the convex hull of \(\operatorname{ES}(\HH_k)\), and it follows that \(\conv{\QQ_k} \subsetneq \conv{\on{GS}(\HH_{2k})}\).

\subsection{B. Proof of Observation 1}

{\bf Observation 1.}
{\it Consider a pure state \(\ket{\psi}\) and a mixed state 
  \(\varrho \geq \delta\eins\) with \(R_k(\varrho)= R_k(\dyad{\psi})\). 
  Then for any state \(\sigma\) in the ball \(\mc{B}_\delta(\ket{\psi})\) 
  a valid state \(\tilde{\varrho}\) in \(\mc{B}_\delta(\varrho)\) can be 
  found, such that their \(k\)-party reduced density matrices match.
  Moreover, the entropy of $\tilde\vr$ is larger than or equal to the entropy of $\sigma$.
  This implies that the ball \(\mc{B}_\delta(\ket{\psi})\) contains no
  thermal states of $k$-body Hamiltonians.
}

\begin{proof}
Any \(\sigma\) in the trace ball \(\mc{B}_\delta(\ket{\psi})\) can be written as 
\(\sigma = \dyad{\psi} + X\), with a traceless $X$. The trace can be
decomposed as 
\(
  \tr(X) = \bra{\psi} X \ket{\psi} + \sum_i \expval{X}{\psi_i^\perp} = 0 \,,
\)
where the $\ket{\psi_i^\perp}$ are orthogonal to $\ket{\psi}.$
The second term of this expression is positive, since 
\begin{align}
  \sum_i \expval{X}{\psi_i^\perp} 
  &= \sum_i \expval{ (X + \dyad{\psi}) }{\psi_i^\perp}
\nonumber  \\
  &= \sum_i\expval{\sigma}{\psi_i^\perp} \geq 0 \,.
\end{align}
So we must have \(\expval{X}{\psi} \leq 0 \). Furthermore, $X$ can
only have one negative eigenvalue $\lambda_-$, otherwise there would be
also $\ket{\psi_i^\perp}$ with $\expval{X}{\psi_i^\perp} < 0$, which is 
in contradiction to $\sigma \geq 0.$ From $\tr(X)=0$ it follows that 
$\lambda_-$ has the largest modulus of all eigenvalues and consequently
\( \tr(|X|) = 2 |\lambda_{-}|\). Since
\( D_{\tr}(\dyad{\psi}, \sigma) = \tr|X| /2 \leq \delta, \) 
it follows that $|\lambda_-| \leq \delta$.
    
For \(\sigma \in \mc{B}_\delta(\ket{\psi})\) we choose \(\tilde{\varrho} = \varrho + X\) 
as a candidate having the \(k\)-RDMs of \(\sigma\). We have
\begin{equation}
  R_k(\sigma) = R_k(\dyad{\psi} + X) = R_k(\varrho + X) = R_k(\tilde{\varrho}) \,.
\end{equation}
Furthermore, \(\tilde{\varrho}\) is a positive semidefinite density 
matrix, because of \(\tilde{\varrho} = \varrho + X \geq (\delta - 
|\lambda_-|)\eins \geq 0 \). Thus, for any state \(\sigma\) 
in \(\mc{B}_\delta(\ket{\psi})\) there exists a state \(\tilde{\varrho}\) in \(\mc{B}_\delta(\varrho)\), such that the \(k\)-RDMs of \(\sigma\) and \(\tilde{\varrho}\) match.

\begin{figure}[t]
  \includegraphics[width=0.3 \textwidth]{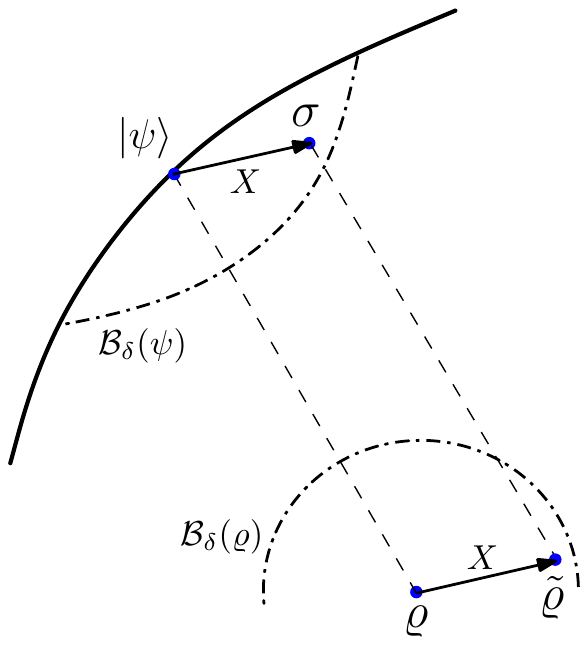}
  \caption{Illustration of Observation 1: If a strictly positive $\vr$
  can be found, then for a given perturbation $\sigma$ of $\ket{\psi}$ one can
  find a corresponding $\tilde\vr$ in the vicinity of $\vr$, such that the reduced
  density matrices of $\sigma$ and $\tilde\vr$ are the same. 
  }
  \label{fig:ball}
\end{figure}

Now we show that the entropy of $\tilde\varrho$ is larger than or equal to the entropy of 
$\sigma$, as this ensures that \(\sigma\) is not in \(\QQ_k\). Namely, 
if the entropy of \(\tilde\varrho\) is larger, a state with the same \(k\)-RDMs but of higher entropy than \(\sigma\) has been found, and \(\sigma\) is outside of \(\QQ_k\). If on the other hand equality holds, then again \(\sigma \notin \QQ_k\) due to the uniqueness of the information projection and because of \(\sigma \neq \tilde\varrho\).

First, note that if \(\rho\) fulfills the condition \(S(\varrho) \geq 2 C_\delta\), where 
\(C_\delta = -\delta \log (\tfrac{\delta}{D-1}) - (1-\delta) \log(1-\delta) \), then also 
as required \(S(\tilde{\varrho}) \geq S(\sigma)\). 
This follows from the sharp 
Fannes-Audenaert inequality \cite{Audenaert2007}
\begin{equation}
  |S(\eta) - S(\mu)| \leq -d \log \big(\frac{d}{D-1}\big) - (1-d) \log (1- d) \,,
\end{equation} 
where \(d = D_{\tr}(\eta, \mu)\) and $D=2^N$ is the dimension of the system.
Recall that \(\sigma \in \mc{B}_\delta(\psi)\) and \(\tilde\varrho \in \mc{B}_\delta(\varrho)\).
Thus the entropy of \(\sigma\) can be at most \(C_\delta\), 
and the entropy of \(\tilde\varrho\) must be at least \(S(\varrho) - C_\delta\). 
Requiring \(S(\varrho) \geq 2 C_\delta\) therefore ensures that the entropy of 
\(\tilde\varrho\) is higher than or equal to that of \(\sigma\).

It remains to show that $\vr$ indeed fulfills this condition. For that, note
that the eigenvalues of $\vr$ are larger than $\delta$ but smaller than $1/N$
due to the normalization of $\vr$. Furthermore assume $D\geq 8$, since 
we are considering at least three qubits. From the bounds on the eigenvalues 
it follows that the entropy of $\varrho$ is bounded by 
\begin{align}
S(\vr)& \geq - [1-(D-1)\delta]\log[1-(D-1)\delta] 
\nonumber \\
&- (D-1)\delta \log(\delta)
\equiv \Gamma \,.
\end{align}
So, we consider the function 
\(
\FF(\delta, D) = \Gamma - 2 C_{\delta}
\)
and have to show its positivity. Let us first fix $D$. Taking the second 
derivative of $\FF$ with respect to $\delta$ one directly finds that this
second derivative is strictly negative. This implies that $\FF$ assumes only
one maximum in the interval $[0, 1/D]$ and that the minima are assumed at the 
borders. We have $\FF(0,D)=0$ and it remains to prove that
$\GG(D)=\FF(1/D,D)$ is positive. For $D=8$ one can directly check that $\GG$ 
as well as its derivative is positive. Furthermore, the second derivative 
of $\GG(D)$ with respect to $D$ is strictly positive for any $D \geq 8$, which 
proves the claim.
\end{proof}

\subsection{C. Numerical results}

Let us first consider states of five and six qubits. We report in Table~\ref{fig:area_numbers}
numerical results for the fraction of pure states lying outside 
of \(\conv{\QQ_2}\), with the condition of positive definiteness \(\delta\) 
ranging from \(10^{-3}\) to \(10^{-7}\). 
We tested \(300'000\) (\(30'000\)) random five-qubit (six-qubit) states distributed
to the Haar measure \cite{Mezzadri07} with our semidefinite program using the solver MOSEK \cite{num:sdp}. 
As can be seen from the Table, 
at least \(40\%\) of all tested five-qubit states and 
\(100\%\) of all tested six-qubit states lie outside of \(\conv{\QQ_2}\). 
Thus, a similar result as in Ref.~\cite{Linden2002} does not hold in the cases of five and six qubits. 

Concerning \(\conv{\QQ_3}\), a single five-qubit state and no six-qubit state has 
been detected to lie outside.
We ascribe the latter result to a rather weak statistics, as
states in the vicinity of \(\ket{M_6}\) are easily 
detected by our semi-definite program (cf. Fig.~\ref{fig:ringcluster5}).

Let us now turn to the case of four qubits. Here, none of \(8\) million random 
pure states have been found to be outside of \(\conv{\QQ_2}\). The numerical 
result suggests that this is a general feature of four-qubit systems. We also 
tested special examples of highly entangled four-qubit states, such as the 
cluster state, classes of hypergraph states \cite{Otfried}, the Higuchi-Sudbery 
\(\ket{M_4}\) state \cite{Higuchi} or the $\ket{\chi}$-state \cite{Guehne09, Osterloh}. 
While many of theses states can be shown to be outside of \(\QQ_2\), we were
not able to prove analytically or with the help of the semidefinite program
that they have a finite distance to \(\QQ_2\). This implies that they might 
be approximated by thermal states of two-body Hamiltonians. 

\begin{table}[t]
\centering
\renewcommand{\arraystretch}{1.3}
\begin{tabular}{|c | *{4}{c|} r|}
\hline
\(\delta\)        & $10^{-3}$ & \(10^{-4}\)     & \(10^{-5}\) &  \(10^{-6}\)   &  \(10^{-7}\) \\
\hline
\(5\)qb  &  0.0040   &  0.1325        &  0.2976     &  0.3729         &  0.4000
\\
\(6\)qb  &  0.7680    & 0.8872    &  0.8897     &  1.0000        &  1.0000\\
\hline
\end{tabular}
\caption{
Fraction of pure five and six qubit states which are outside of the
convex hulls of \(\QQ_2\), as detected by the semidefinite
program from Eq.~(\ref{sdptest}).
See the text for further details.
}
\label{fig:area_numbers}
\end{table}

\subsection{D. Proof of Observation 2}

{\bf Observation 2.}
{\it 
  The maximum overlap between the $N$-qubit ring cluster state 
  \(\ket{C_N}\) and an \(N\)-qubit state \(\tau \in \QQ_2\) 
  is bounded by
  \begin{equation} 
    \sup_{\tau \in \QQ_2} \expval{\tau}{C_N} = \sup_{H \in \mc{H}_2} \tr\big[\frac{e^H}{\tr[e^H]} \dyad{C_N} \big] \leq \frac{D-1}{D},
  \end{equation}
   where $D=2^N$ is the dimension of the system. More generally, for an arbitrary pure 
  state with maximally mixed reduced $k$-party states in a $d^{\otimes{N}}$-system, 
  the overlap with $\QQ_k$ is bounded by $(d^N-1)/d^N$. 
}

\begin{proof}
We consider first only the ring cluster state, the generalization is then straightforward. 
For \(N\geq5\), the ring cluster state \(\ket{C_N}\) has 
\(m(\ket{C_N}) = 3\), that is, all  the two-body reduced density 
matrices are maximally mixed \cite{VandenNest2008}. Since 
the family of thermal states is invariant under the addition 
of the identity \(\tau(H) \mapsto  \tau(H + \theta \eins)\), 
we can choose \(H\) to be traceless when maximizing the overlap.  
So 
$
\tr[H] = 0
$
and 
$\tr[H \dyad{C_N}] = 0$ 
follows. Note that this was the only part in the proof where the property of \(\ket{C_N}\) 
having maximally mixed 2-RDMs was required.

We write \(H\) and \(\dyad{C_N}\) in the eigenbasis \(\{ \ket{\eta_i}\}\) of 
\(H\),
\begin{align}
H &= \sum_i \eta_i \dyad{\eta_i} \nonumber \\
\dyad{C_N} &= \sum_{ij} c_i c_j \ketbra{\eta_i}{\eta_j} \,,
\end{align}
and obtain following conditions, where the second results from the 
normalization of the ring cluster state:
\begin{align}
  f_1  &= \sum_i \eta_i = 0 \,, 			\label{c1}  \\
  f_2  &= \sum_i p_i -1 = 0 \, , && p_i = |c_i|^2\geq 0\,,	\label{c2}  \\
  f_3  &= \sum_i p_i \eta_i = 0				\label{c3} \,.
\end{align}
Under these conditions, we have to maximize
\begin{equation}
\FF = \frac{\sum_i p_i e^{\eta_i}}{\sum_i e^{\eta_i}}.
\end{equation}
If $H$ is nontrivial, it must have both some positive and negative eigenvalues. 
Then at least two of the $p_i$ must be nonzero. We use the 
method of Lagrange multipliers and consider
\begin{equation}
\Lambda = 
\frac{\sum_i p_i e^{\eta_i}}{\sum_i e^{\eta_i}} + \lambda_1 f_1 + \lambda_2 f_2 + \lambda_3 f_3 \,. 
\end{equation}
If the maximum is attained for some value $p_k$ which is not 
at the border of the domain $[0,1]$, we then must have
\begin{equation}
\pd{\Lambda}{p_k} = \frac{e^{\eta_k}}{\sum_i e^{\eta_i}} + \lambda_2 + \lambda_3 \eta_k = 0 \,.   \label{eq:lagrangian1_pk} 
\end{equation}
For a given spectrum of \(H\), \(\{\eta\} = (\eta_1, \dots , \eta_D)\), 
Eq.~\eqref{eq:lagrangian1_pk} has a solution for at most two values, 
\(\eta_+\) and \(\eta_-\).
For any \(\eta_i\) not equal to \(\eta_+\) or \(\eta_-\), the corresponding 
variable \(p_i\) has to lie at the boundary of the domain \([0,1]\), which 
implies that \(p_i = 0\) if \(\eta_i \notin \{\eta_+, \eta_- \}\). The 
eigenvalues \(\eta_+\) and \(\eta_-\) can be \(l\) and \(l'\) fold 
degenerate, with corresponding \(p_+^l, p_-^{l'}\). But then, it is 
easy to see that it is optimal to maximize one of those by taking 
\(p_+ = \sum_l p_+^l\) and \(p_- = \sum_{l'} p_-^{l'}\) and setting 
the others to zero. Second, considering the set of 
$\eta_i \notin \{ \eta_+, \eta_-\}$ where $p_i=0$ one can further
see with Jensen's inequality that it is optimal to take all of 
the $\eta_i$ equal, that is $(D-2)\eta_i = -(\eta_+ + \eta_-)$. 
So, the whole problem reduces to a problem with four variables,
\begin{equation}
  \max_{p_i, \eta_i}
  \FF 
  =
  \max_{p_\pm, \eta_\pm} \frac{p_+ e^{\eta_+}+ p_- e^{\eta_-}}{e^{\eta_+} + e^{\eta_-} + (D-2) e^{-(\eta_+ + \eta_-)/(D-2)}} \,. \label{eq:alpha_sum}
  \end{equation}
From the conditions it follows that we can choose \(\eta_+ > 0\), which implies 
that \(\eta_- = - \eta_+ p_+ / p_- < 0\). We have to prove that the upper bound is
is \((D-1)/D\). Rewriting \(p_- = \frac{\eta_+}{\eta_+-\eta_-}\), 
we aim to show that
  \begin{equation}
  \frac{ \left(1- \frac{\eta_+}{\eta_+ - \eta_-}\right) e^{\eta_+} + \frac{\eta_+}{\eta_+ - \eta_-} e^{\eta_-}}{ e^{\eta_+} + e^{\eta_-} + (D-2) e^{ -(\eta_+ + \eta_-) / (D-2)} } \leq \frac{D-1}{D} \,.
  \end{equation}
This can be rewritten to
  \begin{align}
    &(D-1)(\eta_+ - \eta_-)\big[ e^{\eta_+} + e^{\eta_-} + (D-2)e^{-(\eta_+ + \eta_-)/(D-2)}\big] \nonumber\\
    &- D(\eta_+ e^{\eta_-} - \eta_- e^{\eta_+}) \geq 0 \,.
  \end{align}
Regrouping terms leads to
  \begin{align}
    & \underbrace{(D-1)(D-2) (\eta_+ - \eta_-) \exp \left( -\frac{\eta_+ + \eta_-}{D-2} \right)}_{t_1} \nonumber \\
    & \underbrace{- \left[\eta_+ + (D-1)\eta_- \right] \exp(\eta_-)}_{t_2}  			\nonumber \\
    & \underbrace{+ \left[\eta_- + (D-1)\eta_+ \right] \exp(\eta_+)}_{t_3} \geq 0 		\,.
  \end{align}
The term \(t_1\) is always positive, while the signs of \(t_2\) and \(t_3\) 
depend upon the choice of \(\eta_+\) and \(\eta_-\). So consider the 
following three cases:
  \begin{enumerate}
  \item Case: \( (D-1)\eta_+ < |\eta_-|\):
  Then \(t_2 \geq 0\), but \(t_3 < 0\). However, we have \(t_1 + t_3 \geq 0\) because of
    \begin{equation}
      -(\eta_+ + \eta_-) = -\eta_+ + |\eta_-| \geq  (D-2) \eta_+ 
    \end{equation}
    and
    \begin{align}
       (D-1)&(D-2)(\eta_+ - \eta_-) \nonumber\\
      \geq \, & (D-1)(D-2)|\eta_-|  \geq 2|\eta_-| \nonumber \\
      \geq \, & |\eta_-| + (D-1) \eta_+ \geq |\eta_- + (D-1) \eta_+| \, .
    \end{align}
  \item Case: \( (D-1)^{-1} \eta_+ \leq |\eta_-| \leq  (D-1) \eta_+ \):
  This case directly leads to $t_2 \geq 0$ and $ t_3 \geq 0$.
\item Case: \(  |\eta_-| < (D-1)^{-1}\eta_+\): 
Then \(t_3 \geq 0\), but \(t_2 < 0\). However, we have 
\(t_2 + t_3 \geq 0\), because of \(e^{\eta_+} > e^{\eta_-}\) and
  \begin{align}
    & (D-1) \eta_+ + \eta_- \geq 3 \eta_+ + \eta_- 	\nonumber \\
    \geq \, & 2 \eta_+ \geq \eta_+ + (D-1) \eta_- \, .
  \end{align}
  \end{enumerate}
This finishes the proof. 
  \end{proof}
  
\begin{figure}[t]
\includegraphics[width=0.3 \columnwidth]{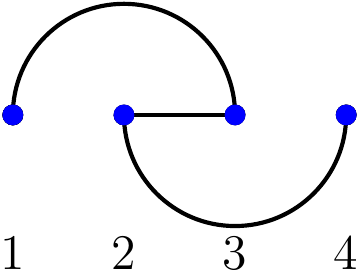}
\caption{Graph of the linear cluster state \(\eta\)
with particles \(2\) and \(3\) exchanged. This state cannot be approximated 
by Hamiltonians with nearest-neighbor interactions only.}
  \label{fig:lin_cluster_perm23}
  \end{figure}

Note that the minimal fidelity distance fidelity distance from 
the convex hull \(\conv{\QQ_k}\) can be used to show the presence of irreducible correlations. 
The minimal distance \(D_k\) to \(\QQ_k\) in terms of the relative entropy is bounded by 
\begin{equation}
D_k(\varrho) \geq \min_{\sigma\in \conv{\QQ_2}} S(\varrho||\sigma) \geq - \log \max_{\sigma \in \QQ_2} F(\varrho, \sigma) \,.
\end{equation}
  This follows from a recent result on \(\alpha\)-Rényi relative entropies \cite{Tomamichel2013},
  \begin{equation}
  S(\varrho||\sigma) \geq S_{1/2}(\varrho||\sigma) = - \log F(\varrho, \sigma) \,.
  \end{equation}
 Therefore, the divergence of the five qubit ring cluster state from 
  \(\QQ_2\) is bounded by \(D_2(\ket{C_5}) \geq 0.0317\).

\subsection{E. Interaction certification}

To certify that higher than two-body interactions have been engineered, 
a four qubit state can be used by further restricting the possible 
interaction structure of the system.
As an example, consider an ion chain of four qubits in a linear trap, 
where the only two-body interactions allowed are of the nearest-neighbor type. 
Then the four-qubit linear cluster state \(\eta\), 
which is a usual linear cluster state with a permutation of particles \(2\) 
and \(3\) (see Fig.~\ref{fig:lin_cluster_perm23}), cannot be obtained as a 
ground or thermal state but only be approximated up to a fidelity of 
\(\alpha = (N-1)/N = 15/16 = 93.75\%\). This value is within reach of
current technologies.

To see why this state cannot be obtained, note that it has the generator 
\(
 G = \{ XIZI, IXZZ, ZZXI, IZIX\}
\),
where \(X, Y, Z, I\) stand for the Pauli matrices and the identity respectively. 
The stabilizer is then given by 
\begin{align}
  S = \{&IIII, IXZZ, IYZY, IZIX, \nonumber\\
        &XIZI, XXIZ, XYIY, XZZX, \nonumber\\
        &YIYX, YXXY, -YYXZ, YZYI, \nonumber\\
	&ZIXX, -ZXYY, ZYYZ, ZZXI \} \,.
\end{align}
The nearest-neighbor marginals of the graph state 
\begin{equation}
\eta = 2^{-4} \sum_{s_i \in S} s_i \,,
\end{equation}
which are \(\eta_{12},\eta_{23}\), and \(\eta_{34}\), are all maximally mixed. 
The remaining two-party marginals do not need to be considered, 
as long-range interactions are precluded by the physical setup.
Then an argument similar to as in Observation \(2\) can be made.
It is again interesting to see what fraction of states cannot be ground states in such a setup.
Our semidefinite program shows that \(94\%\) of pure states cannot be 
approximated as ground or thermal states of a linear spin chain having nearest-neighbor interactions only \cite{qsremark2}.
However, when including next-to-nearest neighbor interactions, 
no unobtainable states were detected.


\begin{thebibliography}{99}
\bibitem{h4note}
The reason is that in the proof of 
Observation 2 one has not only the constraint $\tr[H \dyad{C_{5 \times 5}}]= 0$, 
but also $\tr[H^2 \dyad{C_{5 \times 5}}]= 0$.
\bibitem{qsremark2}
We tested \(3\) million random pure states for \(\delta = 10^{-7}\).
\end{thebibliography}
\end{document}